# An analytical model of tornado generation

S.N. Artekha (С.Н. Артеха)

AFFILIATIONS

Space Research Institute of RAS, Moscow 117997, Russia

Author to whom correspondence should be addressed:  sergey.arteha@gmail.com

## ABSTRACT

A new analytical model for the generation of axisymmetric tornado-type vortices has been developed. A solution to the nonlinear equation for the stream function in an unstable stratified atmosphere is obtained and analyzed within the framework of ideal hydrodynamics. The solution is sought by smooth connecting continuous solutions for the internal region ("eye"), the central region ("wall" with maximum velocities) and the external region of the tornado. Expressions describing radial dependences for the radial and vertical velocity components include combinations of Bessel functions. The vortex is spatially localized by radius and height. Convective instability of a stratified atmosphere leads to an increase in the radial and vertical components of velocities according to the hyperbolic sine law. A downward flow is observed near the tornado axis. The maximum speed of the upward flow is achieved at a certain radial distance at a certain height. Below this height, radial flows converge towards the central part of the tornado, and above this height there is an outflow from the "wall" to the axis and to the periphery. The radial structure of the azimuthal velocity is determined by the structure of the initial disturbance and can change with height. Maximum rotation is achieved in the tornado "wall" at a certain height. The increase in azimuthal velocity can occur according to a superexponential law. Possible structures of movements, scenarios for the development of a tornado and its dynamics are discussed.

## I. INTRODUCTION

Atmospheric vortex structures of different scales significantly influence terrestrial weather and climate. In the variety of atmospheric vortex movements, one can distinguish mesoscale (hundreds and thousands of kilometers in size) and concentrated vortices, which have a noticeable impact on our lives. Concentrated vortices are spatially localized non-stationary vertically elongated vortex structures with a characteristic transverse scale from several meters to several hundred meters (sometimes up to several kilometers). These include dust devils,[3] more intense and large-scale vortices – tornadoes,[18,36] waterspouts,[10] and firespouts, which can occur suddenly during fires in calm weather[59] or during volcanic eruptions.[58] Despite the fact that the listed vortices arise in different environments and can be generated by different natural mechanisms, they all exhibit an upward spiral motion at some distance. The rotation speed reaches its maximum value at a certain distance from the vortex axis and tends to zero at its periphery.

Meteorological observations[51] served as the basis for the creation of the first thermodynamic model of dust devil generation.[46,49] In this model, warm air in a convectively unstable atmosphere rises and then cools and falls. The proposed model is an analogue of a heat engine that draws energy from a hot surface layer. In Ref. 40, a hydrodynamic model of axially symmetric convective vortices (assuming weak disturbances) in a convectively unstable atmosphere at the initial stage of generation was proposed. This model was further developed for various cases of stream function and seed azimuthal velocities (see Refs. 16,41 and references therein). The work[39] presents an analytical model of a spatially limited vortex, applicable for arbitrary values of the coordinates $r$ and $z$. For this purpose, a solution was obtained in the form of Bessel functions (instead of the previously used approximations in the form of power and exponential functions) using the method of searching for stationary large-scale dipole vortices of Rossby waves in the neutral atmosphere.[26]

The most powerful of the concentrated vortices – a tornado – can exceed 4 kilometers in diameter and travel hundreds of kilometers; wind speeds can exceed 500 km/h. Multivortex tornadoes also occur. Unlike other vortices, a tornado always comes into contact with the parent thundercloud at the top.[4,15,36] When it occurs, the trunk of a tornado also descends from above, from the cloud (sometimes a whirlwind begins to rise towards it near the surface, enveloping it from the outside, and they combine into a single tornado). The air inside the funnel descends and the air outside rises, rotating rapidly.

The current research methods include field observations,[5,23,27,63] numerical modeling[28] and physical simulation in the laboratory.[11,29] There are quite a lot of tornado models (analytical, empirical and numerical) and theories: supercell theory, rear-flank downdraft theory, multiple vortex theory, self-organization, electric theory etc. (see Refs. 1,14). Many works have been devoted to the study of tornadoes, touching on the conditions of its formation, thermodynamics, the dynamics of its development, various physical models, and numerical modeling (see, for example, Refs. 4,30,32,65, as well as reviews[12,50] and references therein). Such an important feature of a tornado as a downdraft in a funnel was noted by many authors.[4,7,12,50,61] The review[12] discusses such issues as the rotation of the updraft during a turning shear, the need for the downflow, the baroclinic mechanism, the barotropic mechanism, and the role of friction with the earth's surface. In Ref. 4, the main attention is paid to the role of downdraft and condensation (see also Ref. 30), when expanding, moist air from the cloud is adiabatically cooled in the funnel and condensation begins. As a result of this process, as well as the displacement of droplets due to centrifugal force, the pressure is greatly reduced. A large pressure difference leads to an increase in flow speed, i.e. to intensifying tornadoes. This is a two fluid model. In Ref. 66, a model was built based on the balance of forces and estimates were made that viscosity can be neglected for calculating radial and vertical flow, but viscosity is important for determining the rotation speed.

Despite extensive satellite sensing and supercomputer computing capabilities,[13,33,43,44] only one in five supercells is known to become a tornado. In addition, tornadoes appear not only from supercells, but, for example, from a squall line. The National Weather Service has a false alarm rate of about 70 %. Despite the large number of works and noticeable progress in research, many questions related to the conditions for the formation of tornadoes, their dynamics, the possibility of prediction and influence on them remain open.[33]

Information on the statistics of whirlwinds and tornadoes, as well as some of their characteristics (sometimes there are some terminological disagreements among observers, physicists and meteorologists) can be found both in the scientific literature (see, for example,[8,9,38,57]) and on the websites of specialized organizations in different countries (National Oceanic and Atmospheric Administration, Tornado and Storm Research Organization, European Severe Storms Laboratory, National Center for Atmospheric Research, Federal Service for Hydrometeorology and Environmental Monitoring, etc.) and on popular Internet resources (Wikipedia, Britannica, etc.). Tornadoes come in all shapes and sizes. They are often visible as a funnel of rotating condensate and passive impurity (their actual size exceeds the size of the visible funnel). The width of a tornado at the surface of the earth ranges from 2 m to 4.2 km; the height reaches from 150 m to 1500 m (if you take into account the rotating cloud, it can reach up to 10 km); wind speeds inside the tornado are estimated to be between 100 km/h and 575 km/h; maximum azimuthal velocities near the ground are found at an altitude of 20-60 m; the speed of the tornado as a whole is from 0 to 120 km/h; the distance traveled ranged from 2 m to 352 km; the lifetime of a tornado ranges from several minutes to 3.5 hours; the time it takes for a tornado to form ranges from a few minutes to about an hour (we do not take into account the preparatory stages, for example, the formation of the rotating supercell itself). Average tornado characteristics vary significantly depending on the season and geographic location of occurrence. Due to the wide variety of tornado characteristics, it would be desirable to have a fairly simple but flexible analytical model.

The goals of this work are the following:

– in the approximation of ideal hydrodynamics, to write down an equation for the stream function that describes the developing instability in a stratified atmosphere;

– to obtain rigorous analytical solutions applicable for arbitrary values of the cylindrical coordinates *r* and *z*, which describe the generation of a tornado localized in space;

– within the framework of the constructed analytical model, which includes a number of free parameters, to describe the structure of the tornado velocity field and various scenarios of its dynamics.

The structure of the article is as follows. Section II derives equations for the stream function that can manage the nonlinear internal gravity waves (IGWs). Section III discusses the generation of poloidal motion of tornadoes in an unstable stratified atmosphere. Section IV describes the enhancement of azimuthal rotation in a tornado. Section V discusses the main findings of the study, possibility of generalization of the model and contains conclusions.

**II. BASIC EQUATIONS**

We will consider the axisymmetric case. We introduce a cylindrical coordinate system $(r, \varphi, z)$. The *z* axis is directed vertically upward. We assume that all quantities do not depend on the angle of the cylindrical system $\varphi$, i.e. $\partial / \partial \varphi = 0$. The values of pressure and density, respectively, will be determined by the sum of the equilibrium and perturbed values:

$$p = p_0(z) + \tilde{p}(t,r,z), \quad \rho = \rho_0(z) + \tilde{\rho}(t,r,z), \tag{1}$$

where from the condition of hydrostatic equilibrium

$$\frac{d p_0}{d z} = -\rho_0(z) g, \tag{2}$$

the altitude dependence of the equilibrium pressure is determined:

$$p_0(z) = p_0 \exp\left(-\frac{z}{H}\right), \quad H = \frac{c_s^2}{\gamma_a g}, \quad c_s = \sqrt{\frac{\gamma_a p_0}{\rho_0}}, \tag{3}$$

here $\mathbf{g} = -g\hat{\mathbf{z}}$ is the gravitational acceleration, $\hat{\mathbf{z}}$ is the unit vector along the vertical axis, $\gamma_a$ is the adiabatic exponent (the ratio of specific heat capacities $c_P / c_V$), $H$ is the local scale of atmospheric height, and $c_s$ is the speed of sound.

In the general case, the flow velocity, considered as perturbed motion $\mathbf{v} = (v_r, v_\varphi, v_z)$, can be decomposed into its poloidal component $\mathbf{v}_p$ and azimuthal component $v_\varphi \hat{\mathbf{e}}_\varphi$, i.e. $\mathbf{v} = \mathbf{v}_p + v_\varphi \hat{\mathbf{e}}_\varphi$, where $\hat{\mathbf{e}}_\varphi$ is the corresponding unit vector. The original system of equations in a cylindrical coordinate system includes the following equations:
1) continuity equation for perturbed motion:

$$\frac{\partial \rho}{\partial t} = -\nabla \cdot (\rho \mathbf{v}) = -\rho (\nabla \cdot \mathbf{v}) - (\mathbf{v} \cdot \nabla) \rho. \quad (4)$$

We consider the emerging (disturbed) flow to be incompressible:

$$\frac{1}{r} \frac{\partial (r v_r)}{\partial r} + \frac{\partial v_z}{\partial z} = 0, \quad (5)$$

then the distribution of the perturbed density will be determined by the remaining terms of Eq. (4).

2) Navier-Stokes equations for incompressible flow:

$$\frac{\partial v_r}{\partial t} + v_r \frac{\partial v_r}{\partial r} + v_z \frac{\partial v_r}{\partial z} - \frac{v_\varphi^2}{r} = -\frac{1}{\rho} \frac{\partial p}{\partial r} + \nu \left( \frac{1}{r} \frac{\partial}{\partial r} \left( r \frac{\partial v_r}{\partial r} \right) + \frac{\partial^2 v_r}{\partial z^2} - \frac{v_r}{r^2} \right), \quad (6)$$

$$\frac{\partial v_z}{\partial t} + v_r \frac{\partial v_z}{\partial r} + v_z \frac{\partial v_z}{\partial z} = -\frac{1}{\rho} \frac{\partial p}{\partial z} + \nu \left( \frac{1}{r} \frac{\partial}{\partial r} \left( r \frac{\partial v_z}{\partial r} \right) + \frac{\partial^2 v_z}{\partial z^2} \right), \quad (7)$$

$$\frac{\partial v_\varphi}{\partial t} + v_r \frac{\partial v_\varphi}{\partial r} + v_z \frac{\partial v_\varphi}{\partial z} + \frac{v_r v_\varphi}{r} = \nu \left( \frac{1}{r} \frac{\partial}{\partial r} \left( r \frac{\partial v_\varphi}{\partial r} \right) + \frac{\partial^2 v_\varphi}{\partial z^2} - \frac{v_\varphi}{r^2} \right). \quad (8)$$

3) Equation of state of an ideal gas (Mendeleev-Clapeyron equation):

$$\frac{p}{\rho T} = const. \quad (9)$$

4) We are looking for a solution that is applicable to the initial stage of vortex generation. At this stage, the potential energy reserves of the unstable atmosphere are large, the process of transition to kinetic energy is quite intense, and friction forces play an insignificant role. Besides, in order to be able to obtain an analytical solution, simplifying assumptions should be made. Let us first neglect the dissipative effects of thermal conductivity and viscosity (in Eqs. (6)-(8) we set $\nu = 0$). Then the last one will be the transport equation for potential temperature $\theta = p^{1/\gamma_a} / \rho$, which is a single-valued function of entropy:

$$\frac{d\theta}{dt} = 0. \quad (10)$$

The initial basis for describing convective cells in the atmosphere was thermodynamic models of the generation of vertical currents.[46,47,49] It is now generally accepted that the generation of vertical currents is due to the instability of the stratified atmosphere. Square of the Brunt–Väisälä frequency or buoyancy frequency

$$\omega_g^2 = g \left( \frac{\gamma_a - 1}{\gamma_a H} + \frac{1}{T} \frac{dT}{dz} \right) \quad (11)$$

characterizes IGW. The atmosphere is considered to be unstable stratified if this square is negative. In expression (11), $T$ and $dT/dz$ are the atmospheric temperature and the temperature gradient in the vertical direction, respectively. Due to solar heating of the surface, the vertical temperature gradient (the second term of the Brunt–Väisälä frequency in (11)) is negative, and a situation can arise when its value exceeds the first term. This condition corresponds to the well-known Schwarzschild criterion for convective instability. In this case, disturbances are not carried away from the region of their origin by IGWs. As a result, in the instability region, IGWs turn into unstable, exponentially growing structures.

Equations describing the dynamics of IGWs were previously derived either for the velocity potential or for the stream function in many works (see, for example, Refs. 35,45,54,55). When deriving the basic equation, we will use the stream function in cylindrical coordinates, following the procedure developed in Refs. 40,42,49.

Poloidal velocity components are related to the stream function $\psi(t,r,\varphi,z)$ by the following relations:

$$v_r = -\frac{1}{r}\frac{\partial \psi}{\partial z}, \quad v_z = \frac{1}{r}\frac{\partial \psi}{\partial r}. \tag{12}$$

System of equations (5)–(7), (9), (10) leads, according to Refs. 40,42 to the following equation for the stream function (12), which describes the evolution of nonlinear internal gravity waves:

$$\left(\frac{\partial^2}{\partial t^2} + \omega_g^2\right)\Delta^*\psi + \frac{1}{r}\frac{\partial}{\partial t}J(\psi,\Delta^*\psi) = 0. \tag{13}$$

where $J(a,b) = \frac{\partial a}{\partial r}\frac{\partial b}{\partial z} - \frac{\partial a}{\partial z}\frac{\partial b}{\partial r}$ is the Jacobian determinant and the operator $\Delta^*$ is defined as

$$\Delta^* = r\frac{\partial}{\partial r}\left(\frac{1}{r}\frac{\partial}{\partial r}\right). \tag{14}$$

The Jacobian determinant in Eq. (13) corresponds to the so-called vector nonlinearity.

If $\omega_g^2 < 0$, Eq. (13) describes the nonlinear dynamics of structures in an unstable stratified atmosphere. We will consider the case when instability occurs at the moment $t = 0$, i.e. in (11) we have $\omega_g^2 \to -|\omega_g|^2$. In the opposite case, instability does not arise and the energy of disturbances is carried away from the region of their occurrence with the help of IGWs.

**III. GENERATION OF POLOIDAL MOTION**

Equation (13) can be satisfied if the last term vanishes due to a specific coordinate dependence, and the remaining term vanishes due to a time dependence. Hence, we can look for a scalar stream function that can generate poloidal velocity components using a separation of variables method. Next, for the time dependence of the solution, we obtain two functions $\sim \exp(\gamma t)$ and $\sim \exp(-\gamma t)$, from which we construct the dependence so that the instability manifests itself at the initial moment of time $t = 0$. Therefore, we use the form:

$$\psi(t, r, z) = v_0 r^2 f(z/L) \operatorname{sh}(\gamma t) \Psi(R), \tag{15}$$

where $v_0 = const$ is some characteristic poloidal velocity; The instability increment is determined from (11): $\gamma = |\omega_g|$; $L = const$ is the characteristic vertical spatial scale of the tornado, such that $L \ll H$; $\Psi$ is a function depending on the dimensionless radial distance $R = r/r_0$, $r_0$ is the characteristic radius of the vortex; The function $f(z/L)$ is still completely arbitrary. The function must satisfy the conditions of regularity of the three components of velocity and pressure on the axis of symmetry of the vortex. Substituting the stream function (15) into Eq. (13), we obtain the following equation:

$$J(\psi, \Delta^*\psi) = 0. \tag{16}$$

The nonlinearity of Eqs. (13) and (16) leads to the fact that the solutions (their radial part) are not additive in the radial coordinate. All particular solutions of the nonlinear equation (16) coincide with solutions of the linear equation of the form

$$\Delta^*\psi = A\psi, \tag{17}$$

where the value $A$ is constant.

Radial velocity (12) should vanish at $r \to 0$ and $r \to \infty$, and vertical velocity (12) should vanish at the boundaries of the tornado, i.e. at $r \to \infty$ and $z = 0, L$. Consequently, the desired stream function (15) must remain localized in the radial direction, therefore it must satisfy the conditions:

$$\left(\psi, \frac{\partial \psi}{\partial r}\right) \to 0, \tag{18}$$

when $r \to 0$ and $r \to \infty$, that is, the function must be regular along the axis of symmetry of the cylinder and vanish at infinity. To find a solution to Eq. (17) that satisfies these boundary conditions, we apply the operator $\Delta^*$ to the stream function given by expression (15):

$$\Delta^*\psi = v_0 f(z/L) \operatorname{sh}(\gamma t) \left( R^2 \frac{d^2 \Psi}{dR^2} + 3R \frac{d\Psi}{dR} \right). \tag{19}$$

By choosing $A = \pm a_0^2 / r_0^2$ in Eq. (17), where $a_0$ is a real number, and using (19), we obtain the following linear differential equation for the function $\Psi$:

$$R^2 \frac{d^2 \Psi}{dR^2} + 3R \frac{d\Psi}{dR} = \pm a_0^2 R^2 \Psi. \tag{20}$$

The general solution to the above equation can be represented in the form of several functions, including Bessel functions:

(i) With a negative sign on the right in (20), the solution is an arbitrary combination of functions $\Psi(R) = J_1(a_0 R)/R$ and $\Psi(R) = Y_1(a_0 R)/R$.

(ii) With a positive sign on the right in (20), the solution can be the following real functions: $\Psi(R) = I_1(a_0 R)/R$ or $\Psi(R) = K_1(a_0 R)/R$.

(iii) For a zero value of $a_0$ we obtain the solution $\Psi(R) = C_0 + C'/R^2$.

Since the original equation (13) and Eq. (16) are nonlinear, the sum of the above solutions is no longer a solution, since the constants in Eq. (20) on the right will be different. As a result, it is easy to verify that Eq. (13) will be satisfied identically to each of the above functions (i)–(iii), but the ranges of applicability of these solutions with respect to the variable $R$ must differ. Moreover, we see that with such a choice the function $f(z/L)$ can be chosen completely arbitrarily.

From the found piecewise exact solutions (i)–(iii), it is necessary to combine a whole exact solution that is real and does not give any singularities for the velocity components. For the near-axial region of a tornado, the solution should lead to a zero radial velocity component on the axis. At a characteristic radial distance, a maximum of the radial (and azimuthal) velocity component should be observed. For large distances from the tornado axis, the solution should not oscillate, but rather quickly decrease. The physical requirement for the continuity of the physical quantities themselves leads to the analogous mathematical requirement: solutions for different areas, including velocity components, must fit together smoothly and continuously. Also, the solution should be similar to a real tornado in the structure of movements and in the relationships of all quantities. For example, radial movements in the paraxial region and in the central region can be opposite.

To satisfy conditions (18), we look for a solution to Eq. (20) by combining continuous solutions for the internal region $\Psi_{int}(r < r_1)$ – the "eye" region of the tornado, the central region of the tornado $\Psi_{centr}(r_1 < r < r_2)$, where the radial and azimuthal velocities reach maximums, and for the external region $\Psi_{ext}(r > r_2)$ (periphery of the tornado). The quantities $r_1$ and $r_2$, where the matching of solutions occurs will be determined later. At each boundary between the regions (at $r = r_1$ and $r = r_2$) the corresponding continuity conditions are satisfied:

$$\left(\Psi(r_1), \frac{\partial \Psi(r_1)}{\partial r}\right)_{int} = \left(\Psi(r_1), \frac{\partial \Psi(r_1)}{\partial r}\right)_{centr}, \quad \left(\Psi(r_2), \frac{\partial \Psi(r_2)}{\partial r}\right)_{centr} = \left(\Psi(r_2), \frac{\partial \Psi(r_2)}{\partial r}\right)_{ext}. \quad (21)$$

In the internal region, all the previously listed physical conditions are satisfied by solution (i) of the following form:

$$\Psi_{int}(R) = -C_0 J_1(a_0 R)/R, \quad (22)$$

where the constants $C_0$ and $a_0$ will be defined later.

In addition, we note that near the tornado axis (and only) we could also use the solution (iii); this specific solution leads to a constant vertical component of velocity and linear dependence of the radial velocity component from *R*. However, this solution does not provide any fundamentally new opportunities: as a result, it would have to be used not three areas by *R*, but four, and to join this solution (iii) again with the solution (22) at some additional intermediate point. Therefore, we will straightaway use the solution near the axis (22), which has a much wider range of opportunities (including those close to solution (iii) near the axis).

In order for the solution to include the "eye" of tornado (22), the neighboring central region must be described by the following solution (i):

$$\Psi_{centr}(R) = C_1 \frac{J_1(a_1 R)}{R} + C_2 \frac{Y_1(a_1 R)}{R}, \qquad (23)$$

where the constants $C_1, C_2$ and $a_1$ will be defined later.

In the external region of the tornado, the form of a decreasing solution from (ii) is also one-valued:

$$\Psi_{ext}(R) = C_3 \frac{K_1(a_2 R)}{R}, \qquad (24)$$

where $C_3$ and $a_2$ are the constants, which we will define later.

It is more convenient to obtain all mathematical solutions in dimensionless variables (it is also more convenient to present results graphically in this form), then they have a more universal character (each physical quantity will be made dimensionless, i.e. written as a ratio to some characteristic dimensional value of the same physical nature: $R = r/r_0$, $Z = z/L$, etc.). The fact is that all characteristics of whirlwinds and tornadoes vary over very wide ranges of values. The ranges of characteristic values and their typical values is presented in Section 5. We substitute the solutions we found (22)–(24) into the chosen stream function (15) and write down in the general form of expression for the radial and vertical components of velocity from (12). Then, for the radial component of velocity in the internal $(0 \leq r < r_1)$, central $(r_1 \leq r < r_2)$ and the external region $(r_2 \leq r < \infty)$, we get the following expressions respectively:

$$v_r^{int} = v_0 \frac{r_0}{L} f'(Z) \operatorname{sh}(\gamma t) C_0 J_1(a_0 R), \qquad (25)$$

$$v_r^{centr} = -v_0 \frac{r_0}{L} f'(Z) \operatorname{sh}(\gamma t) [C_1 J_1(a_1 R) + C_2 Y_1(a_1 R)], \qquad (26)$$

$$v_r^{ext} = -v_0 \frac{r_0}{L} f'(Z) \operatorname{sh}(\gamma t) C_3 K_1(a_2 R). \qquad (27)$$

Similarly, for the vertical component of velocity in the internal $(0 \leq r < r_1)$, central $(r_1 \leq r < r_2)$ and external region $(r_2 \leq r < \infty)$, we obtain the following expressions respectively:

$$v_z^{int} = -v_0 f(Z) \sh(\gamma t) C_0 a_0 J_0(a_0 R), \qquad (28)$$

$$v_z^{centr} = v_0 f(Z) \sh(\gamma t) a_1 [C_1 J_0(a_1 R) + C_2 Y_0(a_1 R)], \qquad (29)$$

$$v_z^{ext} = -v_0 f(Z) \sh(\gamma t) C_3 a_2 K_0(a_2 R). \qquad (30)$$

Now our task is to determine all the values of the parameters and coefficients. The ratio of parameters $a_1$ and $a_0$ determines the radial structure and the ratio of the maximum velocities in the central and internal regions. The parameter $a_2$ determines the rate of decay for velocity at the periphery of the tornado. In principle, we can arbitrarily choose the parameters $a_0$, $a_1$ and $a_2$, but then the dimensionless variable $R$ (related to the radius $r_0$) will not be able to determine the characteristic size of the tornado structure. Therefore, 1) we will define the desired characteristic radius $r_1 < r_0 < r_2$ ($R = 1$) as such the radial distance at which the radial velocity is maximum in absolute value (the azimuthal velocity will also be maximum); 2) for uniform comparison of constants, we also renormalize the radial part of the radial velocity (the expression in square brackets in (26)) so that the maximum is equal to unity.

The algorithm is as follows.

1) First, we select arbitrary quantities $a_{00}$, $a_{10}$ of interest to us; let, for example, it be 9 and 2.5 respectively.

2) In order for the values of the function $\Psi$, radial and vertical velocities to match at a certain boundary $R_{10} = r_{10} / r_0$, we solve numerically (we solved it in the Wolfram Mathematica system using FindRoot) the following system of equations for the values $C_{10}, C_{20}, R_{10}$:

$$\begin{cases} J_1(a_{00} R_{10}) = C_{10} J_1(a_{10} R_{10}) + C_{20} Y_1(a_{10} R_{10}), \\ a_{00} J_0(a_{00} R_{10}) = a_{10} (C_{10} J_0(a_{10} R_{10}) + C_{20} Y_0(a_{10} R_{10})), \\ a_{00}^2 J_1(a_{00} R_{10}) = a_{10}^2 (C_{10} J_1(a_{10} R_{10}) + C_{20} Y_1(a_{10} R_{10})). \end{cases} \qquad (31)$$

3) Taking into account the found values $C_{10}, C_{20}$, we are looking for such a distance $R = R_{max}$, at which the value of the radial velocity in the central region is maximum (we equate the derivative with respect to $R$ to zero), i.e. solve the equation for $R$:

$$C_{10} (J_0(a_{10} R) - J_2(a_{10} R)) + C_{20} (Y_0(a_{10} R) - Y_2(a_{10} R)) = 0. \qquad (32)$$

4) We find the value of the maximum radial velocity:

$$k_0 = C_{10} J_1(a_{10} R_{max}) + C_{20} Y_1(a_{10} R_{max}). \qquad (33)$$

5) Now the required quantities are given by the following formulas:

$$a_0 = a_{00} R_{max}, \; a_1 = a_{10} R_{max}, \; C_0 = 1/k_0, \; C_1 = C_{10}/k_0, \; C_2 = C_{20}/k_0, \; R_1 = R_{10}/k_0. \tag{34}$$

We present all the resulting values accurate to 4 decimal places:

$$a_0 = 9.2228, \; a_1 = 2.5619, \; C_0 = 0.9611, \; C_1 = 1.6936, \; C_2 = 1.0724, \; R_1 = 0.4155. \tag{35}$$

6) Now, so that the values of the function $\Psi$, radial and vertical velocities are matched at the second boundary $R_2 = r_2 / r_0$, we solve numerically the following system of equations for the values $C_3, a_2, R_2$:

$$\begin{cases} C_3 K_1(a_2 R_2) = C_1 J_1(a_1 R_2) + C_2 Y_1(a_1 R_2), \\ a_2 C_3 K_0(a_2 R_2) = a_1 \left( C_1 J_0(a_1 R_2) + C_2 Y_0(a_1 R_2) \right), \\ a_2^2 C_3 K_1(a_2 R_2) = a_1^2 \left( C_1 J_1(a_1 R_2) + C_2 Y_1(a_1 R_2) \right). \end{cases} \tag{36}$$

We chose the following solution (you can also specify $R_2$):

$$a_2 = 4.0266, \; C_3 = 337.4226, \; R_2 = 1.5. \tag{37}$$

Using the values of parameters (35) and (37) found above, we can demonstrate the spatial dependence of the radial component of the stream function (15) and its "smoothness" on both boundaries $R_1 = r_1 / r_0$ and $R_2 = r_2 / r_0$ (see Fig. 1). The first boundary corresponds to the intersection $\Psi(R) = 0$, and the second one is shown by the dotted line.

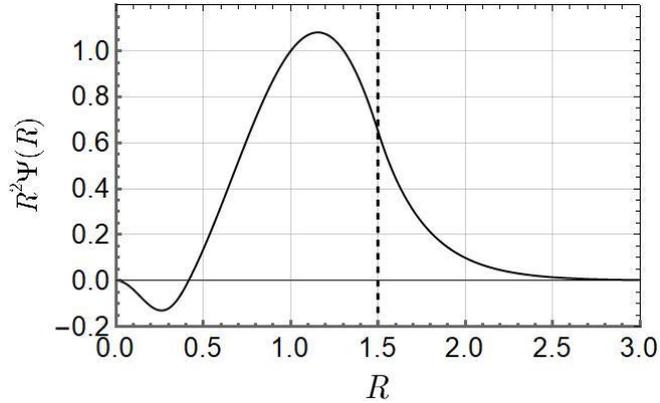

**FIG. 1**. Variations of the radial part of the stream function $R^2 \Psi(R)$ depending on the dimensionless radial distance $R$. Here the connecting boundaries are: $R_1 = 0.4155, \; R_2 = 1.5$.

By choosing the type of function $f(Z)$ of the dimensionless parameter $Z = z/L$, it is possible to achieve different z-dependence of velocity components $v_r$ and $v_z$, for example, so that some component (or both) is turned to zero at the vortex boundaries, reach a maximum or minimum at certain heights.

The dependence of the radial component of the velocity $v_r$ (in units of $v_0$) on the dimensionless distance $R$ from the tornado axis is shown in Fig. 2 for three different times of time; we chose $f'(Z)r_0/L = 0.1$. This picture demonstrates the localization of the flow in the radial direction. It can be seen from the graph that the function found and its derivatives in $R$ are continuous at the boundaries of the regions. Also, the component $v_r$ is regular on the axis of symmetry and turns to zero on the periphery of tornado. The radial velocity converges on both sides to the boundary $R = R_1$, reaches the maximal value (in modulo) at the radial distance $R = 1$ (another extremum is approximately at $R = R_1/2$). At the initial moment, this velocity component is zero, and over time, the growth of the radial component tends to an exponential law (at the saturation stage, viscous forces will limit the increase in speed). Since the function $f(Z)$ should first increase from 0 with an increase in height $Z$ (speed on the surface is zero), and when approaching the height of $Z = L$, the function $f(Z)$ should again decrease to 0, then the radial velocity components at small and high altitudes should be different in the sign (outflow from the axis - inflow to the axis).

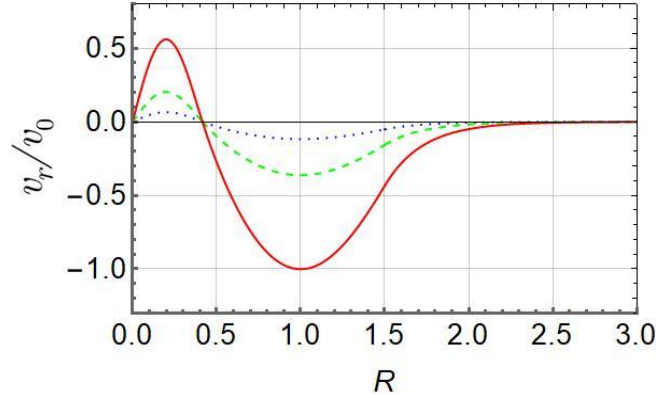

**FIG. 2.** The dependence of the dimensionless function $v_r(R,Z)/v_0$ on the dimensionless radial distance $R = r/r_0$. Dotted, dashed and solid lines correspond the values of $\gamma t = 1, 2, 3$ respectively; such a dimensionless height $Z = z/L$ is chosen that $f'(Z)r_0/L = 0.1$. Here the connecting boundaries are $R_1 = r_1/r_0 = 0.4155$, $R_2 = r_2/r_0 = 1.5$. The maximum inflow corresponds to $R = 1$, the maximum outflow corresponds to $R = 0.2077$, and we normalized the Z-maximum value $v_r(1,Z)/v_0 = 1$.

A similar dependence of the vertical component of velocity (in units of $v_0$) on the dimensionless radial distance from the tornado axis $R = r/r_0$ is shown in Fig. 3 for three different values of $\gamma t$. At the initial moment, the velocity component is zero, and over time, the growth of the vertical component tends to an exponential law (viscous forces will eventually limit the increase in speed and lead to a quasi-stationary stage). This figure also confirms the localization of the flow in the radial direction. It can be seen that $v_z/v_0$ reaches a minimum value in the center of the stream (downward flow). At $R \approx 0.3$, the vertical component of velocity becomes zero. Further, with increasing distance from the axis, at $R = R_1$ the vertical component reaches a maximum. Then, at $R \approx 1.2$, the vertical component of the velocity again becomes zero. Further, in the region $R > 1.2$, the upward flow in the central region of the tornado turns into a downward movement at the periphery with a local minimum of $v_z/v_0$ at $R = R_2$. Further, the z-component of velocity tends to zero at the boundary of the vortex. The magnitude of the z-component of velocity first increases with height from 0, reaches a maximum, and then decreases with height.

The emerging structure of the poloidal movement $v_r$, $v_z$ of the air masses in convective cells describes the vertical flows (rising and descending) growing in time.

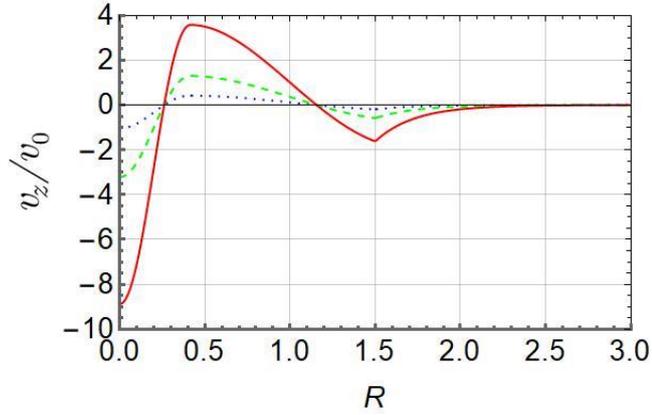

**FIG. 3.** The dependence of the dimensionless vertical component of velocity $v_z(R,Z)/v_0$ on the dimensionless radial distance $R = r/r_0$. The picture corresponds to the choice $f(z/L) = 0.1$. Dotted, dashed and solid lines correspond to $\gamma t = 1, 2, 3$. Here the connecting boundaries are $R_1 = r_1/r_0 = 0.4155$, $R_2 = r_2/r_0 = 1.5$. The maximum of the downdraft is at the axis of the tornado. The maximum of the updraft corresponds to $R_1 = 0.4155$. We see a downward flow for $R < 0.3$ and $R > 1.2$. For $0.3 < R < 1.2$ an upward flow is observed.

### IV. DESCRIPTION OF THE AMPLIFICATION OF THE AZIMUTHAL MOTION

To describe the growth of vortex motion, we use the remaining azimuthal component of the momentum equation (8), taking into account the fact that $\partial/\partial\varphi = 0$ and viscosity $v = 0$:

$$\frac{\partial v_\varphi}{\partial t} + \frac{v_r}{r}\frac{\partial}{\partial r}(rv_\varphi) + v_z \frac{\partial v_\varphi}{\partial z} = 0, \qquad (38)$$

where the radial and vertical components of the velocity are given by expressions (25)–(30).

Since all other quantities can be represented as a product of functions with separable variables, it is most natural that for determining the time evolution and spatial dependence of the azimuthal velocity component, we will also seek a solution for the azimuthal velocity component by the separation of variables method:

$$v_\varphi = y(t) f_0(z/L) V_{\varphi r}(R), \tag{39}$$

where the initial rotation (it is specified by the coefficients $v_{\varphi 0i} = const$) is included in the definition of the value $V_{\varphi r}(R)$. We will again need to find solutions for each individual region in $R$, and then join them at the boundaries. However, for azimuthal velocity this is easier to do, since equation (38) is linear in $v_\varphi$. In order for the solution of Eq. (38) to be determined by such a function with separable variables, we obtain the following system of equations:

$$\frac{d\,y(t)}{dt} = \gamma c_0 \,\mathrm{sh}(\gamma t) \, y(t), \tag{40}$$

$$\frac{\tilde{v}_r}{Rr_0} + \frac{\tilde{v}_r}{r_0 V_{\varphi r}(R)} \frac{dV_{\varphi r}(R)}{dR} + \frac{\tilde{v}_z}{L f_0(Z)} \frac{df_0(Z)}{dZ} = -\gamma c_0, \tag{41}$$

where $c_0 = const$ is a certain number (dimensionless constant), and each component with the tilde icon means the factor of the same function (25)–(30), but without temporary dependence. Then the solution of Eq. (40) will be the following time function:

$$y(t) = \exp\{c_0 (\mathrm{ch}(\gamma t) - 1)\}. \tag{42}$$

Let us now substitute solutions (25)–(30) without time dependence for each of the regions (internal, central, external) into Eq. (41). As a result, for each area we have a similar equation:

$$f'(Z) \left( \frac{\widehat{V}_r(R)}{R} + \frac{\widehat{V}_r(R)}{V_{\varphi r}(R)} \frac{dV_{\varphi r}(R)}{dR} + \frac{f(Z) f_0'(Z) \widehat{V}_z(R)}{f'(Z) f_0(Z)} \right) = -\frac{\gamma c_0 L}{v_0}, \tag{43}$$

where the velocity components with the arch are: $\widehat{V}_r(R) = \tilde{v}_r / r_0$, $\widehat{V}_z(R) = \tilde{v}_z / L$.

From Eq. (43), it is clear that the solution by the method of separating variables in $R$ and $Z$ is possible only if $f'(Z) = const$. In this case, the most general will be the following linear relationship:

$$f(Z) = \begin{bmatrix} k_1 Z, & 0 \leq Z \leq Z_1; \\ \dfrac{k_1 Z_1}{1 - Z_1}(1 - Z), & Z_1 < Z \leq 1, \end{bmatrix} \tag{44}$$

where $Z_1$ is the height where the maximum is reached. In this case, the most general functional dependence for $f_0(Z)$ will be a power dependence:

$$f_0(Z) = \begin{bmatrix} k_2 Z^n, & 0 \leq Z \leq Z_1; \\ \dfrac{k_2 Z_1^n}{(1 - Z_1)^n}(1 - Z)^n, & Z_1 < Z \leq 1, \end{bmatrix} \tag{45}$$

$n > 0$ – any positive real number. Then the solution for the remaining radial dependence of the azimuthal velocity component can be written in quadratures. Accordingly, for the internal $(0 \leq r < r_1)$, central $(r_1 \leq r < r_2)$ and external $(r_2 \leq r < \infty)$ regions we obtain:

$$V_{\varphi r}^{int}(R) = v_{0\varphi 0} \exp\left\{\int_1^R \frac{-\alpha_{1,2}^0 x - nx\widehat{V}_z^{int}(x) - \widehat{V}_r^{int}(x)}{x\widehat{V}_r^{int}(x)} dx\right\}, \tag{46}$$

$$V_{\varphi r}^{centr}(R) = v_{0\varphi 1} \exp\left\{\int_1^R \frac{-\alpha_{1,2} x - nx\widehat{V}_z^{centr}(x) - \widehat{V}_r^{centr}(x)}{x\widehat{V}_r^{centr}(x)} dx\right\}, \tag{47}$$

$$V_{\varphi r}^{ext}(R) = C_4 v_{0\varphi 1} \exp\left\{\int_1^R \frac{-\alpha_{1,2} x - nx\widehat{V}_z^{ext}(x) - \widehat{V}_r^{ext}(x)}{x\widehat{V}_r^{ext}(x)} dx\right\}, \tag{48}$$

where the constants $\alpha_1$ and $\alpha_2$ refer respectively to the lower part $0 \leq Z \leq Z_1$ and to the upper part $Z_1 < Z \leq 1$ along the height of the vortex, and for these areas the following substitutions $c_0 = \frac{\alpha_1 v_0}{\gamma L}$ and $c_0 = -\frac{\alpha_2 v_0}{\gamma L}$ were made. Here the signs are chosen in such a way that at the initial moment of time the seed azimuthal velocity is continuous in height. Opposite signs would mean a gradual attenuation of rotation. Different values of constants $\alpha_1$ and $\alpha_2$ could correspond to different differential rotations and different dynamics of vortex motion in height in the region of radial inflow and outflow. In the viscous case, the friction between two moving layers leads to creation of some transition layer with continuous intermediate velocity components and can be taken into account by introducing an appropriate intermediate function $f_{tr}(Z)$. Naturally, for the continuity of the flow in the horizontal plane, the values of $\alpha_1$ and $\alpha_2$ should be the same in the central region of the tornado $(r_1 \leq r < r_2)$ and the outer region $(r_2 \leq r < \infty)$ of the vortex (which is taken into account in (47), (48)). Another choice would be to rotate with a gap (with a shear). Since the joining of the inner and central regions occurs at the boundary $R = R_1$ at $v_r = 0$, the movements in these regions are essentially separated from each other, and $\alpha$ can differ, which is indicated by the superscript 0 in expression (46). The characteristic azimuthal velocities $v_{0\varphi 0}$, $v_{0\varphi 1}$ may also differ in (46) and (47), including sign. The coefficient $C_4$ was introduced to ensure continuity of the azimuthal velocity when crossing the boundary of the central and external regions (since Eq. (38) is linear in the azimuthal velocity). Next we will consider the case of a single vortex with continuous rotation $\alpha_1 = \alpha_2 \equiv \alpha_0$.

Note that in terms of height, it is possible to construct a model of a tornado from more than two sections, for example, choose different dependencies for the near-surface base of the vortex, the main part of the tornado, and the upper scattering region (note, that in nature the behavior of a tornado at different heights can also differ). With a larger number of vertical sections, it is possible to organize a system of coupled cyclonic-anticyclonic rotation (which is found for tropical cyclones). In this case, the *R*-distribution of the azimuthal velocity and its dynamics may differ for the upper and lower half of the vortex structure.

For further graphical representation, instead of (44), (45), we will make the simplest choice $f_0(Z) = f(Z)$ with a maximum at the middle of the height:

$$f(z/L) = \begin{cases} (z/L), & 0 \leq z \leq L/2; \\ 1-(z/L), & L/2 < z \leq L. \end{cases} \qquad (49)$$

We will continue the calculations with the previous values of all parameters, choosing for certainty and better visibility: $v_{0\varphi 0} = -0.4v_0$, $v_{0\varphi 1} = v_0$, $\alpha_0 = 0.01$. Although the probability of rotation in different directions (and even increasing) is extremely small, we will consider this particular case here, and will discuss all other scenarios in the next Section.

Figure 4 shows the dependence of the azimuthal velocity component $v_\varphi / v_0$ on the distance *R* from the axis. The localization of rotational motion in the radial direction is visible. By changing the value of $\alpha_0$, different differential rotation can be obtained. The azimuthal velocity reaches maximum values at *R* = 1 (another local maximum will be at approximately $R = R_1/2$). At the maximum in height (at *z* = *L*/2) the velocity will be 5 times greater than that shown on the graph. If the radial and vertical velocities increase by approximately *e* times with each increase in unit $\gamma t = 2 \to 3 \to 4 \to \cdots$ (the growth tends to be exponential), then the growth of the azimuthal velocity in the lower half of the vortex tends to be super-exponential. At first, the azimuthal velocity component increases slowly due to the smallness of $\alpha_0$, but after $\gamma t = 5$ its increase sharply outpaces the growth of the other two velocity components (approximate increase for each unit of *γt*: 1.07, 1.19, 1.6, 3.58, 32.01, 12352.6, … times). Therefore, we did not depict this growth at small *γt* and at large *γt*, but took intermediate time moments.

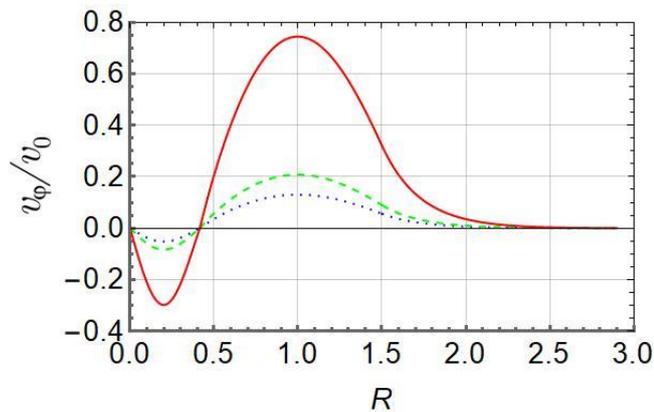

**FIG. 4.** The radial dependence $v_\varphi(R, Z)/v_0$ at height $z/L = 0.1$. Dotted, dashed and solid lines correspond to $\gamma t = 4, 5, 6$. Here the connecting boundaries are: $R_1 = r_1/r_0 = 0.4155$, $R_2 = r_2/r_0 = 1.5$. The maximum rotation speed is observed at $R = 1$, the second extremum corresponds to R = 0.2077.

The detailed behavior of all the spatial properties of the velocity components described above can be seen in Fig. 5, which shows contour graphs of the velocity field on the $R - Z$ plane. The first graph corresponds to the radial velocity component, the second graph corresponds to the vertical component, and the third graph corresponds to the azimuthal velocity component. We see the spatial localization of currents inside a tornado with extrema at certain distances $R$ and height $Z$. The movements disappear at the periphery. Let us remind that the values of all parameters were chosen arbitrarily, only for the convenience of graphical representation.

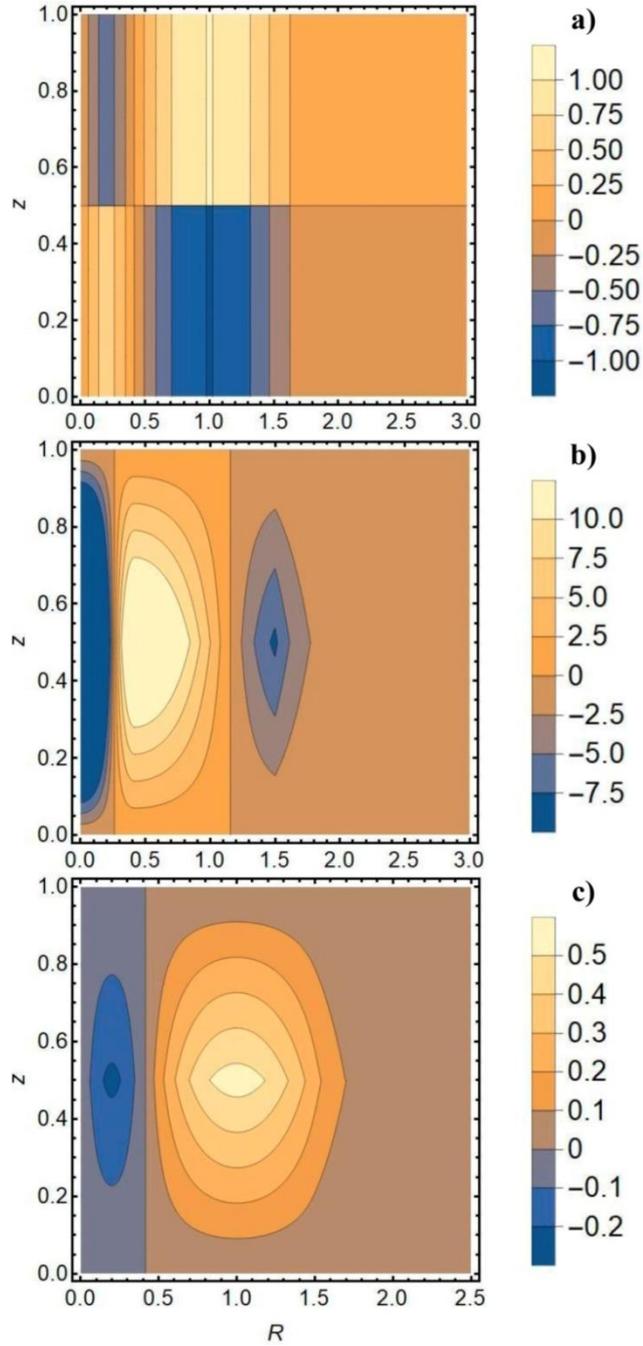

**FIG. 5.** The tornado velocity field at $\gamma t = 3$: a) $v_r(R,Z)/v_0$, extrema correspond to $R = 0.2077$ and $R = 1$; b) $v_z(R,Z)/v_0$, extrema correspond to $R = 0$, $R = 0.4155$ and $R = 1.5$ at $Z = 1/2$; c) $v_\varphi(R,Z)/v_0$, extrema correspond to $R = 0.2077$ and $R = 1$ at $Z = 1/2$.

## V. DISCUSSION OF RESULTS, POSSIBLE GENERALIZATION AND CONCLUSIONS

The proposed tornado generation mechanism, applicable for an unstable stratified atmosphere, is based on the Schwarzschild criterion for convective instability. In this case, the process can proceed from two sides: an overheated near-surface air mass with a superadiabatic temperature gradient (lighter) can begin to rise upward, and a supercooled cloud-air mixture (heavier) can begin to descend downward (towards). The generation model, in energy essence, is a heat engine. Due to the action of the continuity equation, excess energy activates and strengthens wind flows. Of course, thermal convection tends to reduce thermal stratification, but this is a long process, since in place of rising light heated air there is a radial influx of warm air from neighboring areas (similarly, for descending heavy cold air). At the preliminary stage, the strong vertical flow of moist air is very important for tornadogenesis, since it is thanks to such a flow that the cumulonimbus cloud itself is formed and maintained (movements of tornadic air masses affect part of the cumulonimbus cloud also).

In the case of $\omega_g^2 > 0$, convective instability does not arise. In this case, all hyperbolic functions in the derived solutions turn into similar trigonometric functions. As a result, the disturbances are carried away from the region of their origin by the IGWs, and the structure does not arise.

In this work, a nonlinear equation is obtained for the stream function in an unstable stratified atmosphere, which describes the generation and growth of axially symmetric structures. This is one of the possible methods for describing convective instability. A stream function $\Psi(R)$ is proposed, which allows one to reduce this nonlinear equation to an equation that has various Bessel functions as solutions. By matching solutions at the boundaries of convective cells separating different areas of the tornado, an analytical solution was obtained for the radial and vertical velocity components, valid for all distances $R$ and $Z$. The radial structure of the azimuthal velocity is determined by the structure of the initial disturbance. In addition to the instability leading to vertical and radial motion, rotation in ideal hydrodynamics increases and is physically maintained due to the conservation of angular momentum. Therefore, the vortex will grow or be maintained until the entire region with the initial non-zero twist is pulled towards the axis due to the radial inflow (hence, the value of $\alpha_0$ should be very small), and for a moving vortex - until it moves in the instability region with non-zero twist (non-zero helicity). In ideal hydrodynamics, this model describes a vortex localized in space, for which the law of conservation of mass and angular momentum is satisfied. However, in the viscous case, the presence of excess energy can cause the energy, when released, to turn into kinetic energy of rotation and to increase the angular momentum. Let's remember, for example, a funnel in a bathtub.[53] When opening the cork, we can slightly twist the water near the hole in either direction. The evolution of the vortex in the bath is determined by the non-zero initial angular momentum of the fluid and the angular momentum created by viscous forces.

From the obtained solution it is clear that the azimuthal velocity in the proposed model is different from zero (an initial rotation) at $t = 0$. We can calculate the initial angular momentum in a large volume of a cell from which the vortex will concentrate. Then we can calculate the angular momentum of an already formed tornado, and equate these values:

$$\int_{V_{cell}} r\rho(r,z,0)v_\varphi(r,z,0)r\,dr\,dz = \int_{V_{torn}} r\rho(r,z,t)v_\varphi(r,z,t)r\,dr\,dz.$$

From here we can estimate the limit value of $\gamma t$ and, therefore, the upper limit of the flow velocity. To do this, it is first necessary to numerically calculate the density dependence $\rho(r,z,t)$.

The structure of the solution is presented in Fig. 5 (where all parameters were chosen arbitrarily). In the lower part of the vortex, the radial flows converge to the boundary $R = R_1$, while the radial velocity reaches a maximum at $R = R_1/2$, and a minimum at $R = 1$. In the upper part of the vortex, the situation is exactly the opposite (inflows and outflows change places). The vertical velocity component describes a downward flow near the axis with an absolute minimum on the tornado axis. At $0.3 < R < 1.2$, an upward flow is observed with a maximum at $R = R_1$. At $R > 1.2$ there will again be a downward flow with a local minimum at $R = R_2$. The radial and vertical velocity components begin to increase from zero according to the hyperbolic sine law. Vortex rotation turns out to be differential. The radial structure of the azimuthal velocity is determined by the structure of the initial disturbance. The rotation speed increases according to a super exponential law. The rotation will be maximum at $R = R_1/2$ and $R = 1$ at a certain height. From experiments with smoke rings it is known that a rotating torus demonstrates the stability of its motion. In essence, the structure of movements in a tornado that we obtained represents two nested ones along a common boundary, rotating along the same axis of the torus, i.e. must have increased stability.

If the rotation in the inner region and in the central region of the tornado occurs in opposite directions, then it may turn out that the difference in $v_\varphi$ is close to the speed of sound, or even $\Delta v_\varphi > c_s$ – greater than the speed of sound. When such a strong tornado strikes an object, its effect can be with great destructive consequences.

For ideal hydrodynamics, all scenarios are equally probable. Thus, the work examines in detail the case when the rotation in the paraxial region is opposite to the rotation in the main region of the tornado, and both rotations are intensified. But there may be other, more realistic (for real hydrodynamics) scenarios. Let us consider schematically some other cases. Thus, Fig. 6 shows the case of a decreasing rotation in the inner region, when a region of relative calm gradually appears there (for example, due to viscous friction with the neighboring main region of the tornado). Unfortunately, observations of the interior of a tornado are extremely limited.

There is only evidence that there exists often an area of relative calm inside large tornadoes. How it is formed is unknown. Either it forms immediately, along with the tornado itself, or, along with the sharply growing size (and power of the tornado), movements in the inner region begin to fade. Therefore, we will consider all mathematical possibilities.

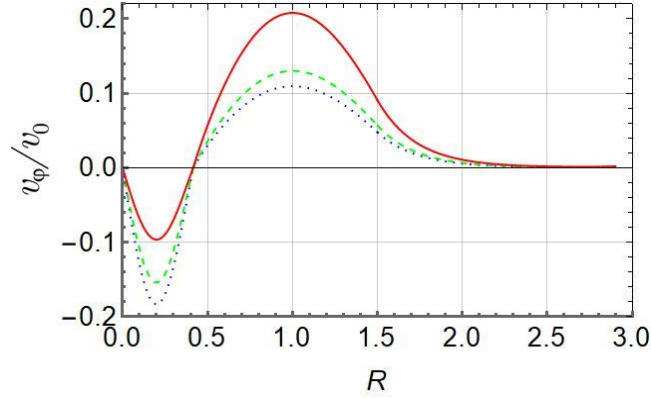

**FIG. 6.** The formation of an area of calm, when $v_\varphi^{int}$ decreases, and $v_\varphi^{centr}$ increases. The dotted, dashed and solid lines correspond to $\gamma t = 4, 5, 6$.

Another scenario is depicted in Fig. 7, when both rotations occur in the same direction and become stronger. When viscosity is taken into account, these two rotating regions will tend to unite into a single rotating vortex with a smoother profile. Similar dependences with two maxima are often observed.[62,64] We can introduce some small addition to the found solution $\mathbf{v} \to \mathbf{v} + \mathbf{v}'$ and substitute it into the Navier-Stokes equations; As a result, the profile is smoothed out and the speed increase slows down (a more rigorous numerical calculation requires solving a complete system of equations).

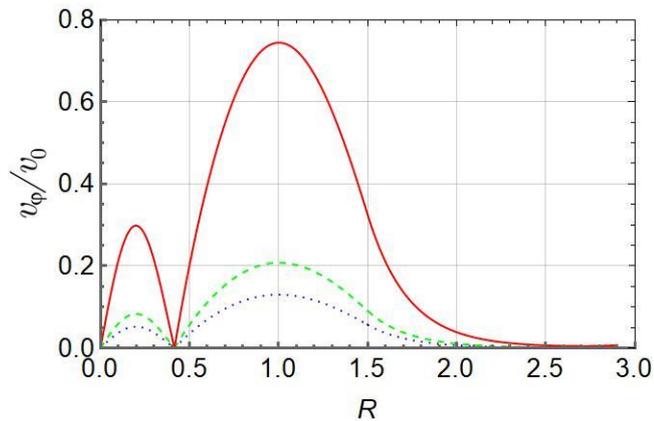

**FIG. 7.** The case of increasing rotations occurring in one direction. Dotted, dashed and solid lines correspond to $\gamma t = 4, 5, 6$.

We will not depict the same case of codirectional rotation, but when the rotation of the internal region fades and gradually a region of relative calm forms in the paraxial region. Also, in all the cases considered above, a scenario can occur when the rotation in the upper part of the vortex fades and tends to simply radial movements.

All previously drawn radial profiles of the azimuthal rotation velocity used the simplest dependence (49). For the more general case of power-law dependence (44), (45), from expressions (46)–(48) one can also draw the azimuthal velocity profile. For $n < 1$, the curve will pass more sharply through the boundary $R = R_1$, and for $n > 1$, the transition across the boundary $R = R_1$ will be smoother, as shown in Fig. 8 for the case $n = 2$.

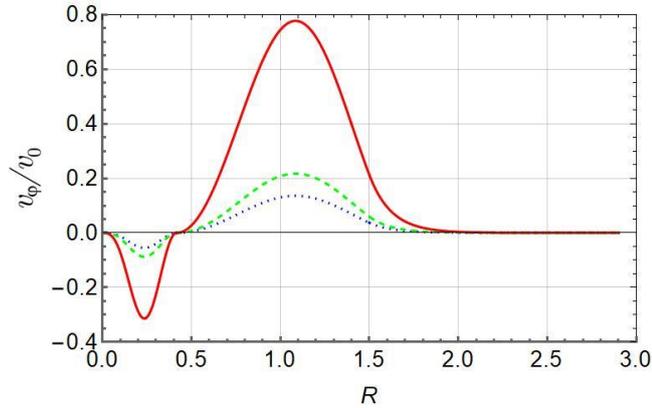

**FIG. 8.** The case of increasing rotations occurring in different directions at $n = 2$. The dotted, dashed and solid lines correspond to $\gamma t = 4, 5, 6$.

Note that the solution for a narrower tornado (without an "eye") can be composed of two regions in $R$ using formulas like (22) and (24), as was done in Ref. 39.

Friction with the earth's surface can be taken into account by introducing a function $f_{surf}(Z)$ such that the velocity components vanish at the surface. In the case of z-layers with different directions of movement, the friction between two mutually moving layers leads to creation of some transition layer with continuous intermediate velocity components and can be taken into account by introducing an appropriate intermediate function $f_{tr}(Z)$. In the case of an arbitrary dependence $f(Z)$, the variables $R$ and $Z$ in Eq. (38) are not separated, and instead of Eq. (43), to find the spatial dependence of the azimuthal velocity, the following equation must be solved numerically in a unified way:

$$f'(Z)\left(\frac{\hat{V}_r(R)}{R} + \frac{\hat{V}_r(R)}{V_\varphi(R,Z)}\frac{\partial V_\varphi(R,Z)}{\partial R} + \frac{f(Z)\hat{V}_z(R)}{f'(Z)V_\varphi(R,Z)}\frac{\partial V_\varphi(R,Z)}{\partial Z}\right) = -\frac{\gamma c_0 L}{v_0},$$

which also does not present any fundamental difficulties. In general, we used the method of separation of variables solely so that it was possible to obtain an analytical solution. Naturally, there may be other solutions, with a different shape and dependencies inside the tornado, which cannot be represented as a product of such functions. They can be obtained numerically (we do not know whether other analytical solutions exist).

In physical essence, the initial unstable profile (2) determines a certain reserve of potential energy, which can turn into the energy of poloidal motion. It can be arbitrary. Mathematically, it is included in the expression of the Brunt–Väisälä frequency and determines the rate of intensification of the generated structure. The fact is that the structure of equation (13) allows any dependence on $z$ to be included in the Brunt–Väisälä frequency $\omega_g(z)$. The form of the solution for the stream function $\psi$ will not change. This only leads to a dependence of $\gamma(z)$ on height $z$, i.e. to the differential dependence of growth rate on height $z$. In the expression for the vertical velocity component, only $\gamma$ will be replaced $\gamma \to \gamma(Z)$, and in the expression for the radial velocity component, in addition to the same replacement $\gamma \to \gamma(Z)$, new expression will appear instead of $f'(Z)$: $f'(Z) \to f'(Z) + f(Z)\gamma'(Z)\mathrm{Coth}\left[\gamma(Z)t\right]$. To find the azimuthal component of the velocity, Eq. (38) needs to be solved numerically, since all variables in Eq. (38) are not separated in this case. Taking into account latent heat affects the temperature field. This influence will change the dependence of the Brunt–Väisälä frequency on the coordinates. If we take into account only the dependence on $z$ (and the dependence on the radius is considered weak), then in the solution for the stream function only the dependence of $\gamma(z)$ on $z$ will be changed, while the changes of vertical and radial velocity components can be found simply (see above). If we also take into account the radial dependence $\gamma(r,z)$, then there will be no separation of variables and it is necessary to solve numerically equation (17), and then Eq. (38).

Since there is a huge variety of characteristics of tornadoes and whirlwinds in nature, it is desirable for tornado models to be flexible. Thus, the proposed analytical model of a tornado in the most general case depends on thirteen parameters: $a_0, a_1, a_2, r_0, L, k_1, k_2, n, v_0, v_{0\varphi 0}, v_{0\varphi 1}, c_0$ and $\gamma$. This makes it possible to vary within wide limits the spatial structure of the tornado, the structure of the velocity field, the values of the maximum and minimum velocity components, their location, the growth time of the tornado and the trends of its development. The first eight parameters allow you to completely vary the spatial structure of the vortex in radius and height. The next three parameters determine the relative magnitudes of the velocity components and the direction of rotation. The last two parameters set the dynamics of tornado development over time.

Let us present the characteristic values of the free parameters of the model in dimensional form. Mean value $r_0 = 50$ m (range 30 m to 1.5 km, since smaller tornadoes are most likely composed of two regions rather than three); Mean value $L = 500$ m (range 150 m to 1.5 km); Mean value $v_0 = 10$ m/s (range 5 m/s to 20 m/s); Mean value $v_{0\varphi 0} = 10$ m/s (range 5 m/s to 20 m/s); Mean value $v_{0\varphi 1} = 20$ m/s (range 15 m/s to 30 m/s); Mean value $\gamma = 0.005$ c$^{-1}$ (range of values $\gamma = 0.001 \div 0.01$ c$^{-1}$. The characteristic range of values of dimensionless parameters is as follows: $a_0 = 3 \div 10$ ; $a_1 = 1 \div 3$ ; $a_2 = 3 \div 5$ ; $k_1, k_2, n = 1/2 \div 2$ ; $c_0 = 0 \div 0.01$. Apparently, the model can be used at times $t \leq 5/\gamma = 1000$ c, after which a transition to the quasi-stationary stage will be observed, and viscosity must be taken into account.

For spatial variables, the solution is applicable until the velocities become equal to the background values. The range of applicability of the proposed model is limited to a relatively thin layer of the atmosphere, where convective instability develops. Physically, the region of applicability of the solution is limited by the region of instability and the region with non-zero torque. The analytical solution obtained in this work describes the generation and initial stage of tornado development. The applicability of the solution is limited in time, since to describe the transition to the quasi-stationary stage and describe such a stage, instead of the equations of ideal hydrodynamics, it would be necessary to use the Navier–Stokes equations. This significantly complicates finding an analytical solution and requires numerical calculations.

At $t > 0$, the pressure field begins to change noticeably in space and time. For the considered stage of the process, the final pressure $p = p_0(z) + \tilde{p}(t, r, z)$ may differ by 5-20 % from the initial pressure $p_0(z)$. The distribution of pressure deviation $\tilde{p}(r, z)$ in a tornado from the equilibrium pressure $p_0(z)$ can be calculated by solving a Poisson-type equation:

$$\frac{\partial^2 \tilde{p}}{\partial r^2} + \frac{1}{r}\frac{\partial \tilde{p}}{\partial r} + \frac{\partial^2 \tilde{p}}{\partial z^2} = \frac{g\mu}{R_g}\frac{\partial}{\partial z}\left(\frac{\tilde{p}}{T}\right) - \frac{\mu(p_0 + \tilde{p})}{R_g T}\frac{1}{r}\frac{\partial}{\partial r}\left(rv_r \frac{\partial v_r}{\partial r}\right),$$

where $\mu$ is the molecular weight and $R_g$ is the universal gas constant.

All solutions obtained in the work are based on the already existing convective instability, but for this a number of quantities had to exceed threshold values. The surface layer temperature associated with excess heat must be high for a super adiabatic temperature gradient to occur

$$\frac{dT}{dz} > \frac{\gamma_a - 1}{\gamma_a}\frac{g\rho_0 T}{P_0},$$

and sizes of the unstable area both in the near-surface layer and in the cloud layer are large enough for a tornado to have time to organize $\min\{\Delta x_{hot}, \Delta y_{hot}\} \gg 2r_0$.

The initial twist plays an important role in this phenomenon. On the one hand, it symmetrizes poloidal movements about a single axis. On the other hand, it breaks the symmetry with respect to the direction of rotation. Wherein the initial twist can be quite weak, since the subsequent intensification of rotation in the real viscous case occurs due to the excess energy of the "heat engine mechanism". Rather, what is more important here is the proximity of the spatial profile of the movement to the solution found. Then these movements are "picked up" by the system and self-consistently amplified. The following may claim to be the cause of the initial rotation: aerography, horizontal temperature gradients (and, as a consequence, pressure gradients), shear flows during the collision of air masses, electro-magneto-hydrodynamics of a thundercloud. Let's discuss some assumptions. The uneven distribution of aerosols contributes to uneven heating of the atmosphere and the earth's surface, the appearance of local pressure gradients, and the generation and intensification of vortices.[17] Along with aerography and collision of air flows, all this can lead to rotation around a vertical axis (although the situation applies more to the near-surface layer of the atmosphere). However, in this case the directions of rotation would be equally probable, while observations indicate that the cyclonic rotation of the tornado is noticeably predominant. Taking into account that a tornado is attached to a thundercloud (usually to a cumulonimbus, but sometimes to a cumulus cloud), and a tornado begins to develop from a descending cloud base (a rotating funnel), the key role is played by the rotation of the section of the thundercloud at a certain height. An important role in the implementation of the cyclonic rotation of a thundercloud can be played by the presence of charged plasma-like subsystems there,[2] i.e. electro-magneto-hydrodynamics of a thundercloud (a factor constantly acting in the required direction). Then cases of anticyclonic rotation may correspond to oppositely charged clouds. The fact is that quite strong electric fields and charged layers have been found in hurricanes (and thunderclouds). In such plasma-like subsystems, many instabilities can be excited, leading to the generation of vortex motions.[6,25,31,34,37,52,60] Rotation, for example, occurs in crossed electric (radial) and magnetic fields. Electric fields and their gradients influence clustered ions, i.e. increase the lifting force and speed of all flows. But these are only hypotheses that need experimental verification. By the way, lightning in the eye wall of a tornado leads to ionization and strong local heating of the gas, which increases the vertical flow, which in turn increases the radial inflow (acts like a pump – due to the continuity of the flow), and due to the conservation of angular momentum, this contributes to an increase in the rotation speed. All these electrical influences can be taken into account by changing the value of $\omega_g(r,z)$. This dependence leads to the need to first solve (most likely numerically) Eq. (17), and then Eq. (38), since the variables are no longer separated. But the most important role in the energy of a tornado is played by latent heat: supercooled water vapor from the cloud descends in the inner region of the tornado, is thrown towards the wall by radial flow (and centrifugal force), and condenses into droplets just in the eye wall. As a result, additional heat is generated, increasing the vertical flow in the eye wall.

Next assumption. Measurements inside a tornado show a noticeable temperature difference between the main part and the axial part of the tornado.[56] Perhaps, in addition to the obvious upward warm flows and downward cold flows, the effect is strongly reminiscent of the Rank–Hilsch vortex effect.[19] In our model we also have two flows rotating in the opposite direction and moving in countercurrent. And a similar effect of separation by temperature. But this is also just a hypothesis. All temperature influences can be taken into account by changing the value of $\omega_g(r,z)$.

There are certain difficulties along the path of a direct quantitative comparison of the proposed analytical model and the real phenomenon or other numerical calculations (Figure 5 is for illustrative purposes only and shows one of possible solutions). Any direct measurements of tornado characteristics are impossible due to the destruction of measuring instruments. There are no direct measurements of the pressure field, density field, temperature field, or velocity field in all the tornado area (all data are estimates). And without such data, quantitative comparison with the real phenomenon is difficult. Some characteristics are assessed from a sufficient distance (further it is only possible to approximately extrapolate the values). The order of magnitude of the velocity is determined remotely using two methods: photogrammetry (estimating the speed of movement of impurity objects in the image plane) and the Doppler effect (estimating the speed of movement of impurity particles to or from the radar/lidar). Even remote detection of tornadoes using various radars often occurs with delays (and misses), not to mention the ability to detect the initial stage of tornado development. And the resolution of radars from such a distance is usually rather low. Doppler radar cannot directly measure areas very close to the ground, but the highest velocities are found in the boundary layer (near-surface layer).[22,24] The Doppler velocity is strictly an average of the particle motion within a radar pulse volume projected along a radar beam, weighted by the returned power from each particle. The central region of a tornado is characterized by the absence of scattering particles (little data). In addition, the data is selective (incomplete), since it is subject to processing and preliminary interpretation based on noise and speeds. To measure the velocity field inside a tornado, it is necessary that at least two high-resolution radars be located fairly close to the tornado passing nearby. There were only a few such cases, or researches used the ground-based velocity track display (GBVTD) technique.[21,22,24,27,48] Thus, the comparison could only be with such a specific individual tornado. Not to mention catching the initial stage of tornado development. Data is available for already formed tornadoes (at the saturation stage due to viscosity), i.e. extrapolation needs to be made. In the best case, any comparisons of the results with reality or with other models could only be made qualitatively, in form or order of magnitude. A comparative analysis of some tornado models is contained in Ref. 20. Numerical calculations also use different hypotheses of tornado formation and assumed (not exactly known) initial and boundary conditions, or unknown random forces and configurations (inside a supercell, for example). It is not yet possible in numerical computations to take into account turbulent and real vortex motion of different scales (in essence, the proposed model describes the organization of large-scale averaged motions, and smaller-scale motions are concentrated in turbulent pulsations outside the framework of this model). In nature, there is a huge variety of tornadoes and whirlwinds, differing in shape, structure, size, speed, growth times and existence times, and any computation is possible only if all specific conditions are specified. Therefore, it is necessary to select a specific individual tornado instance, for which measurement data is available. Tornadoes often exhibit an eye-type structure[5,63] (this can be described by a two-region model, as in Ref. 39), but tornadoes with two "eyes" also occur;[21,63] this structure has also been observed in laboratory simulations.[29] The proposed model can describe such structures. Thus, the distribution of speeds in a formed tornado (in m/s), when cut by a horizontal plane $z$ = constant, is shown in Fig. 9. The same data is used here as for Fig. 8, only $v_{0\varphi 0} = 0.3 v_0$, $r_0 = 400$ m. The maximum speed is 75 m/s. We see the double eye structure. All depicted structures and their characteristics are close to the corresponding values measured in Ref. 21, Fig. 1 (Figure 7 in Ref. 63 has a similar form), only corrected for the violation of axial symmetry due to translational motion.

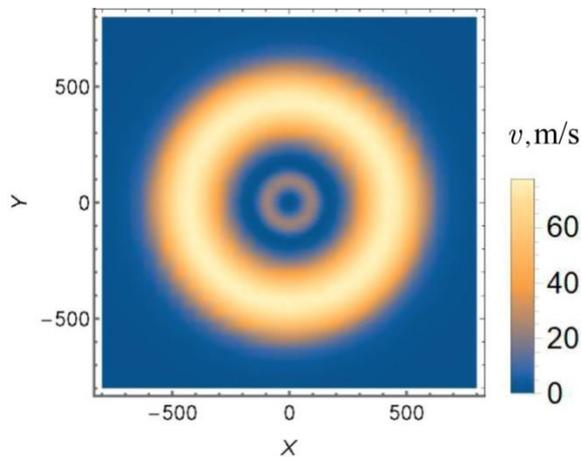

**FIG. 9.** The distribution of speeds in a tornado (in m/s) on the *x-y* plane (in m). The maximum speed is 75 m/s. We see the double eye structure.

Let us take as a basis the scenario shown in Fig. 7, with $v_{0\varphi 0} = 0.3 v_0$, $r_0 = 250$ m. Using the fact that the highest radial and tangential velocities are achieved in the lower part of the vortex,[22,24] and constructing the function $f(z/L)$ so that as $z$ increases, it first increases strongly, then very slowly, then decreases slowly, and then decreases quickly, we obtain the velocity distribution in Fig. 10. The three-dimensional structure of movements and the order of all quantities resembles what is observed for a tornado.[21,23,27]

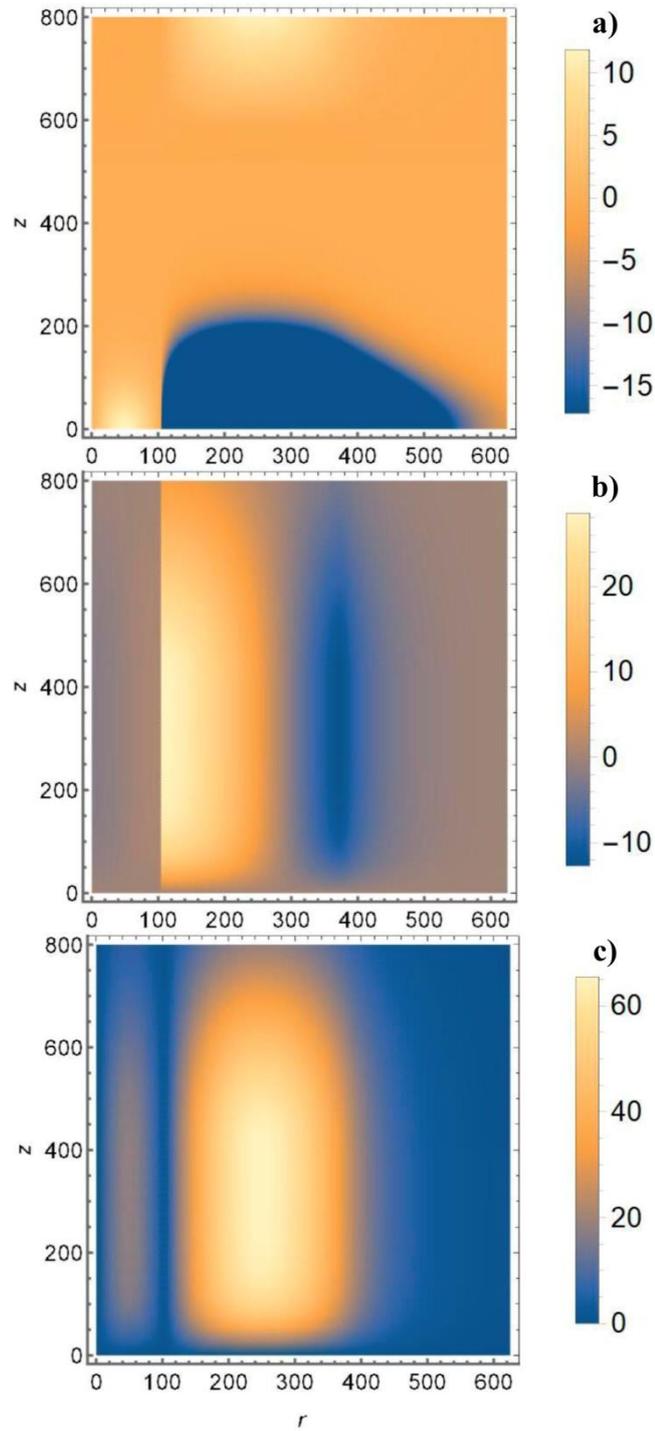

**FIG. 10.** The tornado velocity field in m/s depending on the coordinates $r$ and $z$ in meters: a) $v_r(r,z)$; b) $v_z(r,z)$; c) $v_\varphi(r,z)$.

Perhaps the fine structure of the tornado eye is difficult to measure (or researches have some doubts about the interpretation of multidirectional movements), but researches provide data only for "solid" tornadoes. Such tornadoes can be described using two regions, similar to Ref. 39. If we choose $\delta = 0.08$, then we get $r_1 \approx 1.3121$, $m \approx 1.172114$ (we can choose any values of $\delta = 0.01 \div 0.1$ and find $r_1$ and $m$). In this case we obtain the normalized azimuthal velocity shown in Fig. 11. Similar dependencies measured for many tornadoes (also at different heights) can be seen in many works, for example Refs. 23,27. The results of the proposed model fit neatly among the measured curves. (Among the measured curves there is also a curve of the Burgers-Rott model.[23]) We recall that all comparisons of our model results with observations use extrapolation to the saturation stage. Also, a good agreement can be obtained by comparison with laboratory simulations.[29,48] Thus, the results given by the proposed model are consistent with observational and experimental data.

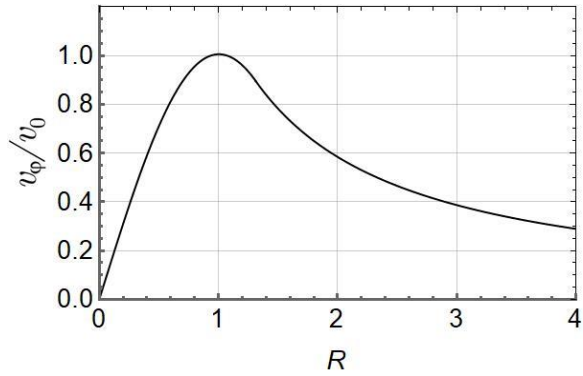

**FIG. 11.** The dependence of the normalized azimuthal velocity on the normalized radius $R$.

We use an ideal axisymmetric model, but in reality many reasons break such symmetry. The movement of the top of the tornado is tied to the movement of the mother thundercloud. In ideal hydrodynamics, with a completely smooth surface, the vertical tornado as a whole would move along with the thundercloud. In reality, it is affected by forces associated with horizontal pressure gradients, friction with a non-uniform surface, including collisions with material objects. However, a tornado has high energy and high angular momentum. It resembles a spinning top (gyroscope), only it is not solid, so it does not bounce off even large obstacles: most of the opposing forces are quite weak for it. However, their multidirectional random actions lead to the complex movement of such a "gyroscope". In addition, the deviation of the earth's surface from the pure horizontal or the axis of the tornado from the vertical leads to a change in the cross-section of the tornado from circular to ellipsoidal. Such shape disturbances propagate upward in a spiral and also affect the complex dynamics of tornado motion. For a strong tornado, such effects are hardly noticeable, but for a weaker and narrower tornado, even multidirectional winds blowing at different heights can lead to bending of the rope (the structure of movements tries to support itself, and the tornado behaves like a bending pump hose). Taking into account the movement of entrained objects denser than air makes the task even more complicated. For them, the force of gravity is approximately compensated by the lifting force of the vertical flow, and the centrifugal force is approximately compensated by the action of the radial flow. As a result, rotating objects visualize a tornado wall - an area of high vertical and azimuthal velocities. The properties and behavior of a non-axisymmetric tornado should be studied only by numerical computations of a complete system of equations, including the condensation of steam with the release of latent heat and the movement of a passive impurity carried away by flows.

We conclude by formulating a summary list related to the presented analytical model. To begin the process of tornado formation, an unstable stratified atmosphere with a presented parent thundercloud and an area with an initial swirl are presumed. The model also includes the initial *z*-elongation of the vortex. We use incompressible flow and exclude viscosity. As a result, the law of conservation of mass and the law of conservation of angular momentum are satisfied for the vortex. Kinetic energy is not conserved, since the potential energy of the unstable atmosphere is converted into kinetic energy of motion. The hydrostatic balance is disrupted in the process of tornado generation. At the generation stage, the cyclostrophic balance is also disrupted.

The main goal of this article was to obtain an analytical solution of the problem. Instead of Eq. (13), we can write a generalized equation for the stream function taking into account viscosity (taking the rotor operation from the Navier-Stokes equation and using other equations, as in Ref. 42). However, the resulting equation will be the rather complicated nonlinear equation of the second order in time and of the fourth order in coordinates. It is difficult to find its analytical solutions. The rigorous numerical calculation is also difficult to carry out, since it is not known which boundary conditions to take for the higher derivatives: there are many options to choose from. This may be the subject of further research.

The tornado phenomenon is quite complex and multifaceted, so it is obviously impossible to cover all the associated nuances of this extensive research topic in one article. An important further direction in the development of the tornado model is the search for quantitative expressions of some introduced free parameters with the real characteristics of an unstable atmosphere and a cumulonimbus cloud. Perhaps statistical analysis could help here.

## ACKNOWLEDGMENTS

The work was carried out within the framework of the State assignment on the topic of fundamental scientific research "Monitoring" IKI RAS (122042500031-8).

## AUTHOR DECLARATIONS

### Conflict of Interest

The author has no conflicts to disclose.

## DATA AVAILABILITY

The data that support the findings of this study are available from the corresponding author upon reasonable request.